\documentstyle{article}
\begin{document}

\title{Fermions, Bosons, Anyons, Boltzmanions\\
        and Lie-Hopf Algebras\thanks{Contribution to: {\it Quantum 
Groups, Deformations and Contractions} -
Bogazici University, Istanbul, Turkey - 17-24 September 1997} }

\author{ E. CELEGHINI ~~$<$celeghini@fi.infn.it$>$\\
Dipartimento di Fisica dell'Universit\`a \\
and I.N.F.N., I50125 Firenze, Italy}

\maketitle

\begin{abstract}
Usual quantum statistics is written in Fock space but it is
not an algebraic theory.
We show that at a deeper level it can be algebraically formalized
defining the different statistics as (multi-mode) coherent states
of the appropriate (but different from the usual ones) Lie-Hopf groups. 
The traditional connection between groups and statistics, established in
vacuum, is indeed subverted by the interaction with the thermal bath.
We show indeed that $h(1)$, related in quantum field
theory to bosons, must be used to define in presence of a bath
the Boltzmann statistics while, to build the Bose statistics, we have
to take into account $su(1,1)$.
Astonishing to describe fermions we are forced to use not the superalgebra 
$h(1|1)$ but $su(2)$ in the fundamental representation.
Higher representations of $su(2)$ allow also 
to give a possible definition of anyon statistics with generalized Pauli 
principle. 
Physical implications are discussed; the results is more general then the
usual on the discrete spectrum, but everything 
collapses to standard theory when the continuum limit is performed. 
\end{abstract}

\section{Introduction}

Group theory provides a natural mathematical language to formulate 
symmetry principles and to derive their consequences in Physics.
The role of symmetries can be of different kind.
We have first kinematical invariances as $su(2)$ for rotations or $o(1,3)$ and
$io(1,3)$ for special relativity. They relate observers and observed systems
denying the existence of a preferential reference frame. Namely they connect
observations of the same system by different observers, stating symmetry
properties of the substratum (like, for instance, homogeneity of the space).
The second kind of physical symmetries is the dynamical invariance that, in 
some sense, is a
self-contradictory concept because, to allow for observations, it must be
broken: without the breaking induced by electromagnetic 
interactions we cannot even see proton and neutron as distinct object.
It express a symmetry of the free states and of their interactions: a
example is the standard model of electroweak interactions,
$su(2) \otimes u(1)$.
A particular case is the $su(3)$ of color symmetry that shows itself in a
very indirect way and would deserve a deeper analysis.
At the end there are (more related to this paper) the second quantization 
algebras:
the two algebras usually considered ($h(1)$ for bosons and $h(1|1)$ for
fermions) are introduced to satisfy physical requirements like 
symmetry or anti-symmetry of states.

These different applications exhibit that connection between
mathematics and physics is not one to one, but quite more complex.
The conclusion is, of course, that group theory is simply a tool allowing 
different applications: the best known example of this fact is $su(2)$, used 
for invariance both under rotation and isospin. 
But, as the same algebra can describe different physical quantities, it is
also possible that the same object fits in different algebraic scheme if the
physical situations is different.
It is exactly what happen in the following.

The different physical situation we consider in this paper is the presence
of a thermal bath.
While in all examples considered before the system was in vacuum, the only
effect of which was in kinematical symmetries, 
the system in this case is not isolated but, on the contrary, controlled by its
interaction with the external word, i.e. a system with infinitely many degree
of freedom. This means that, in absence of interaction,
predictions are essentially predictions on the mean effects of the bath. 

At first sight the enterprise of introducing algebra in the play looks 
impossible:
all the examined applications share the property to use the formalism of
vectorial
linear spaces (i.e., from group theory point of view, representations), while
quantum statistics is not a 
linear theory and, indeed, up to now algebra was not involved in its 
description essentially because we have to take into account not only the 
probabilistic 
effects related to measure in quantum mechanics, but also classical 
probabilistic effects, for instance by means of a density matrix.
But we are able to introduce non-linearity, removing the coherence among
the states, using of the freedom
of introducing in matrix elements arbitrary 
phases that can also be dependent from an external parameter like time.
The expected weights of different vectors are then obtained from coherent
states theory~\cite{Per}, combined with the prescriptions of the theory of
Lie-Hopf algebras to realize multi-mode operators~\cite{F}.  
In same sens we arrive thus to reproduce the situation of kinematical 
invariances:
algebra still describes the properties imposed on the system by the environment
(in kinematics the structure of space, here the effects of the bath).

The main result of our discussion will be that it is possible to reduce usual
postulates of quantum statistics to a deeper and simpler algebraic postulate.

We have also to stress that there are words that
assume different meaning in function of the context. Let us consider indeed
the definition of {\it boson} or
{\it fermion}: in quantum field theory they are simply related to Pauli 
principle while in quantum statistics they are connected to classical 
probability described for instance by the structure of density matrix.
There are not {\it a priori} reasons why the postulates of
quantum statistics and the corresponding definitions used when the system
is truly isolated must imply ones the others.

The scheme of the contributions is this:
we first discuss the case of statistics not bound by the Pauli exclusion
principle, showing that there is room not only for the Bose bosons but also
for the Boltzmann statistics.
Our results show that the connection between algebras and statistics, imposed
by the interaction with the external word, is different from the one obtained 
in quantum field theory. Indeed while, in quantum field theory, $h(1)$ gives 
the algebraic description of bosons, we find that,
to reflect correctly the effects of bath, bosons in quantum statistics
must be related to $su(1,1)$, while $h(1)$ gives the quantum statistics
distribution we associate to Boltzmann statistics. The point is that
both $h(1)$ and $su(1,1)$ allow unlimited occupation 
numbers and the choice between them must be done by the requirement imposed 
by the environment: while microcausality is used to arrive to $h(1)$ as unique
candidate, the weight factors imposed by the 
thermic bath in quantum statistics assign $h(1)$ to Boltzmann statistics and 
$su(1,1)$ to Bose one.

After that, we consider fermions. We might suppose that we have only to extend 
the same procedure and that the toil is easy. 
On the contrary we are forced to change completely the scheme we are used to:
to obtain the correct statistical distribution we have to bring up not
(as expected) the superalgebra $h(1|1)$, but its connected even structure,
$su(2)$ in the fundamental representation $D_{1/2}$ . Work on this unforeseen
development is in progress, attempting to describe our approach in some
sort of bosonization~\cite{MM} or into
supergroup theory~\cite{Cor}.

Next, higher representations of $su(2)$ are considered and it is shown they
are related to generalized Pauli principle. It is a possible new definition of 
anyon statistics~\cite{Hal}. 

In the conclusion the role of continuum limit is considered. As well known, 
but usually not enough stressed, in a limit process a lot of information is 
loss. Usual description of continuum is obtained quantizing in a box and 
pushing the volume to infinity at the end. Thus
while from a discrete theory the continuum is univocally predicted,
experimental results in the continuum (as are most results obtained up to
now in quantum statistics) are not cogent confirmations of a discrete theory.
Using of the power of the algebra, we are indeed able to define
generalized bosons in function of two parameters: $\alpha$, the
coherent state eigenvalue and $k$, the representation highest
weight. The result looks quite new and different from the Bose one, obtained
for $\alpha = 1$ and $k = 1/2$, 
but the limit $V \to \infty$ is independent from $\alpha$ and $k$ and 
traditional predictions on continuum physics are again obtained.

The same thing happens
to our anyons and fermions in the sense that, loosing in the limit the
distinction between contiguous modes, in the continuum the results becomes
independent from the representation of $su(2)$  .
The fundamental difference comes from experimental situation:
while for fermions a lot of physics is obtained with discrete spectrum,
only recent experiments on Bose condensation offer to consider bosons in
the same situation.

Anyway, because whenever algebras have been introduced in some area of 
physics the
impact has been quite sensible, we think that the proposed framework could
be the scheme where quantum statistics will be described in future.

This work is part of a research line under development in cooperation with 
M Rasetti and G Vitiello. On the results about bosons the reader can also 
consult refs. \cite{CRV} and \cite{Bose_cs}. The discussion of fermions and
anyons is new.

\section{Quantum Statistics of Bosons}

A system of identical particles in thermal equilibrium with its environment 
is described in textbooks  (see {\sl e.g.} ref.\cite{HUA}) by a vector in 
Fock space,  
\begin{eqnarray}
   |\psi \rangle = \sum_{\{ n_p \}} c_{\{ n_p \}} \, |n_1,n_2, \dots \rangle 
   \; ,\label{old} 
\end{eqnarray}
and two postulates. The
\underline{\it A Priori Probability Postulate} is a statement on norms
$|c_{\{ n_p \}}|^2$ that are prescribed to be, for all accessible states, 
identically $1$ for Bose and Fermi statistics and  
$1 / {\prod n_p !}$ for Boltzmann statistics.
Note that, in the gran-canonical setting we consider here
(it is the case of phonons in a solid or photons in a black-body),
for bosons the vector $|\psi \rangle$ is not normalizable also for 
finite number of modes.
The \underline{\it Random phases postulate} states that the phases of 
$c_{\{ n_p \}}$'s
~vary quickly and independently in time such that all measurable
interferences are zero. In other words only diagonal
operators play a role, because the time average annihilates all non-diagonal
matrix elements.  

The object of this note is to show that these assumptions can be
derived from algebra, starting from the postulate that the state of 
eq.(\ref{old}) may be defined as (multi-mode) coherent states
of the appropriate Hopf algebras.  

In such a way a connection between statistics and algebras will be established.
Two point must be stressed and will be better discussed in the following: 
first there is not reason why 
the usual correspondence in quantum field theory (that connects bosons with 
$h(1)$, fermions with  $h(1|1)$ and Boltzmann statistics with nothing)
must be saved in this different context; second distributions depend strongly 
from
the representation we work in, but this 
dependence is completely lost in
the continuum limit. 
Because most of bosonic physics,
up to now, has been made in this limit (the new experiment in harmonic traps
at low temperature will be discussed at the end) the dependence from the
boson representation is, for the moment, unphysical.

Let us start to deal with $h(1)$. 
To satisfy the Random Phases Postulate it is sufficient to recall
that an arbitrary phase can
be added to the usual definition of the creation operator with the unique 
requirement
that the opposite phase is added to $a$; 
thus we define $a^\dagger$ such that:
\begin{eqnarray}
a^\dagger |n\rangle = e^{i \chi_n (t)} \sqrt{n+1} |n+1\rangle 
\, .\;\label{a+}
\end{eqnarray} 
{\it I.e.} we change the standard convention, including an independent 
factor of modulus 1 (function 
of $n$ but also quickly varying in function of an external parameter $t$, we 
interpret as the physical time).

Standard coherent states with eigenvalue $1$
are then constructed, following {\it e.g.} ref.~\cite{Per}
\begin{eqnarray}
\phantom{1}\hskip 3truecm e^{a^\dagger}\; |0\rangle \;=\; 
\sum_{n=0}^\infty \;\frac{e^{i \phi_n (t)}}{\sqrt{n!}}
]|n\rangle , \hskip 1.4truecm
\bigl( \phi_n(t) \equiv {\sum_{l=0}}^{n-1}     {\chi}_{l} (t)\bigr) . \nonumber 
\end{eqnarray}
By inspection, the coefficients $c_n$'s are exactly the ones required by 
Boltzmann statistics: we have $|c_{\{ n_p \}}|^2 = 1/n!$ \,and
the presence of quickly and independently varying phases $\phi_n(t)$
guarantees that states with different $n$ are incoherent, as 
required. The only restrictive fact is that the considered Fock space is 
trivial.
Thus we need to extend our approach to a generic Fock space
$\{|n_1,n_2,\dots\rangle\}$:
to build coherent states there, we must start from a multi-mode vacuum 
$|0,0,\dots\rangle$ on which we have 
to 
define a corresponding multi-mode creation
operator. It is at this point that mathematics comes in our aid: as well 
known in Lie-Hopf algebras~\cite{F} the multi-mode 
creation operator on this multi-mode space is simply the iterated
coproduct  $\Delta^{M} (a^\dagger)$ of $a^\dagger$ (where $M$ is the number of 
modes); in physical, less formal, notation simply 
the sum of the corresponding creation operators:
\begin{eqnarray}
\Delta^{M} (a^\dagger) = \sum_{i=1}^M {a_i}^\dagger \;. \nonumber
\end{eqnarray}
Simple calculations give now the coherent state
\begin{eqnarray}
{\rm exp}\bigl[\Delta^M (a^\dagger)\bigr]~~ |0,0,\dots \rangle~ = 
~~\sum_{\{n_i\}}
~\frac{e^{i \phi_{\{n_i\}}(t)}}{\sqrt{n_1! n_2! \dots}}
~~|n_1, n_2 \dots \rangle \, . \label{Bolt}
\end{eqnarray}
Incoherence of states is assured because all phases
$\phi_{\{n_i\}} (t) \equiv  \sum_i  \phi_{i, n_i}(t)$ can be assumed 
independently
and  quickly varying functions of time and the modula ~$|c_{\{n_i\}}|$~ are
such to give the Boltzmann statistics. Thus we have demonstrated our first
statement: Boltzmann particles are coherent states, with eigenvalue 1, of
the Lie-Hopf group $H(1)$.

To change from eq.(\ref{Bolt}) to the Bose's flat distribution we have to 
modify eq.(\ref{a+}). By inspection, all we need is
\begin{eqnarray}
a^{\dagger} |n\rangle = e^{i \chi_n (t)} (n+1) |n+1\rangle 
\label{a+'}
\end{eqnarray}
{\it i.e.} to remove from eq. (\ref{a+}) the square root {\it vinculum}.
With this change, eq. (\ref{Bolt}) indeed becomes
\begin{eqnarray}
{\rm exp}\bigl[\Delta^M (a^\dagger)\bigr]~~ |0,0,\dots \rangle~ = 
~~\sum_{\{n_i\}}
~e^{i \phi_{\{n_i\}}(t)}
~~|n_1, n_2 \dots \rangle , \label{Bose}
\end{eqnarray}
and phases and norms are exactly the required ones to have bosons.

We thus obtain our goal to write 
${|\psi \rangle}_{Bose}$, the vector that
effectively describes a system of bosons, but whit an {\it ad hoc} hypotesis.

The astonishing result is that, by inspection, eq.(\ref{a+'}), with its 
hermitian conjugate for the creation operator $a$ , generates the uirrep 
$D_{1/2}^+$ of $su(1,1)$ (where the Cartan
subalgebra is $H \equiv N+1/2$
and usual notations for $su(1,1)$ can be re-established by
the correspondence $\{ a^\dagger \leftrightarrow J_+, a \leftrightarrow~J_-,
H \leftrightarrow J_3 \}$):
\begin{eqnarray}
[ H, a^\dagger ] = a^\dagger ,\;\;\;\;
~~[ H, a ] = a ,\;\;\;\; [ a^\dagger, a ] = - 2 H .
\nonumber
\end{eqnarray}

From the point of view of group theory the multi-modes representation is
nothing else that the  $M$ times symmetrical
tensorialization of  $D_{1/2}^+$ with highest weight 
$|0,0,\dots \rangle$ and it results to be $D_{M/2}^+$.

Thus we arrive to our second statement: traditional bosons are coherent states
with eigenvalue 1, of the representation $D_{1/2}^+$ of $SU(1,1)$.

We can now appreciate the power of algebraic formalism looking to the offered
opportunities to generalize the usual bosons.
A whole class of two parameters generalized bosons can indeed be defined 
considering not only coherent states with eigenvalue 1 but,
in general, with eigenvalue $\alpha \in {\cal C}$ and not only the 
representation  $D_{1/2}^+$ but the generic representation  $D_k^+$
(where $k \in {\cal I}^+/2$ for $SU(1,1)$, but $k \in {\cal R}^+$ for 
~$\widetilde{SU}(1,1)$, the universal covering group of $SU(1,1)$).

Detailed calculations can be found in ref \cite{Bose_cs}, we quote here only
the results.    
Random phases are obtained always in the same way, while one has, in general 
\begin{eqnarray}
    | c_{\{ n_p \}} |^2 = |\alpha |^{2N}\, \prod_{p} \frac{\Gamma (n_p 
    +2\kappa )}{\Gamma (2\kappa )\Gamma (n_p+1)} \; . \label{ccc}   
\end{eqnarray}

Results look quite different and the reader could suppose that our
bosons have nothing in common with the original ones.
The distributions (\ref{ccc}) vary indeed a lot with $k$ and $\alpha$:
for fixed $N$, $k < 1/2$ enhances for most of
particles in the same state while, for $k = 1/2$, as discussed before we have
the
Bose's bosons and, for $k > 1/2$, a {\it Boltzmann-like} behaviour that becomes
indistinguishable from Boltzmann distribution for $k >> 1$.
On the other hand, $\alpha$ changes the weight of states with different 
values of the total number $N$ of particles: 
if $|\alpha|<1$  ($|\alpha|>1$)  are more (less) weighted states with low $N$
(we must stress that only for $|\alpha|<1$ distributions are normalizable).
This point will be discussed in conclusions.

\section{Quantum Statistics of Fermions}
\indent The same two postulates of quantum statistics of bosons hold for 
fermions too with the obvious restriction imposed by the Pauli exclusion
principle on the allowed states: $n_i \le 1$.
We can thus attempt to verify if our algebraic procedure allows to obtain
the distribution of fermions too. 
The algebra to start with is, of course, $h(1|1)$ we can consider generated by
$\{I; a^\dagger, a\}$:
\begin{eqnarray}
[ I, \bullet \,] = 0\,,\;
~~\{ a, a \} = \{ a^\dagger, a^\dagger\} = 0\,,\; ~~\{ a, a^\dagger\} = I .
\nonumber
\end{eqnarray}
Stated $N \equiv a^\dagger a$\,, the only inequivalent representation can be 
written as

\begin{eqnarray}
a^\dagger = e^{i \chi (t)} \left | \matrix{ 0 & 1 \cr 0 & 0 \cr } \right | 
\;\;\;\;\;
a = e^{- i \chi (t)} \left | \matrix{ 0 & 0 \cr 1 & 0 \cr } \right | 
\;\;\;\;\;
I = \left | \matrix{ 1 & 0 \cr 0 & 1 \cr } \right | \;\;\;\;\;
N = \left | \matrix{ 1 & 0 \cr 0 & 0 \cr } \right | \; ,
\label{matrices} 
\end{eqnarray}
and the standard coherent state~\cite{Per} for one mode with eigenvalue $1$
is then constructed as
\begin{eqnarray}
e^{a^\dagger} \;|0\rangle \;=\; 
|0\rangle \;+\; e^{i \chi (t)} \; |1\rangle .  \label{1f} 
\end{eqnarray}
The result is exactly the required one in a trivial 1-mode Fock space.
We are thus encouraged to attempt the same approach that worked for bosons, 
using the coproduct in Hopf-Lie superalgebras to define, as next step,
to 2-mode coherent states for fermions.
Unfortunately the procedure does not work.
Indeed
we have not problems to build the 2-mode exponential but, because
$\{ a_1^\dagger, a_2^\dagger \} = 0$, we find
\begin{eqnarray}
{\rm exp}~\bigl[ a_1^\dagger + a_2^\dagger \bigr]~~ |0,0\rangle~ \;=\; 
 |0,0\rangle~ +
e^{i \chi_1(t)} \;|1,0\rangle \;+\; 
~e^{i \chi_2(t)} ~\;|0,1\rangle  \label{Fer}
\end{eqnarray}
{\it i.e} vacuum and one-particle states are as required but 
the two-particles state is lost.

The problem is the same for all $M > 1$: only vacuum and one-particle states 
survive.
There is something {\it wrong} ~in the relations among different modes.
Again with our constructive approach, let us look for what relations we need: 
we see that, while the representation of $a^\dagger$ in eq.(\ref{matrices}) 
is satisfactory 
(because gives us the correct one-mode relations), we have to change the
anticommutators  $\{ a_1^\dagger, a_2^\dagger \} = 0$ into commutators
$[ a_1^\dagger, a_2^\dagger ] = 0$.
We have three possibilities (very different for a mathematician, but all 
effective to reach our goal of commuting creation operators): 
we can save our superalgebra $h(1|1)$ and 
introduce same sort of bosonization~\cite{MM},  we can save our 
superalgebra $h(1|1)$ and construct differently the exponential by means
of a related supergroup~\cite{Cor} or we can change the algebraic structure.
Because it is simple and interesting let us consider this last choice:
to build our one-mode exponential (\ref{1f}), all we need are the matrices  
in eqs.(\ref{matrices}); but in these matrices we can read not only $h(1|1)$ 
but $su(2)$ in the fundamental representation $D_{1/2}$ also. 
Explicit calculations show indeed that 
$\{ a^\dagger, a, H \equiv N - 1/2\}$, as defined in (\ref{matrices}), close
$su(2)$:
\begin{eqnarray}
[ H, a^\dagger ] = a^\dagger\,,\;\;
~~[ H, a ] = a\,,\;\; [ a^\dagger, a ] = 2 H .
\nonumber
\end{eqnarray}
Again the coproduct $\Delta^M (a^\dagger)$ defines the multi-modes 
representation  $D_{1/2}^{\otimes M}$ that, on the highest weight 
$|0,0,\dots \rangle$, results to be $D_{M/2}^+$;
the $SU(2)$ coherent state with eigenvalue 1 is thus:
\begin{eqnarray}
{\rm exp}\bigl[\Delta^M (a^\dagger)\bigr]~~ |0,0,\dots \rangle~ = 
~~\sum_{n_i \le 1}
~e^{i \phi_{\{n_i\}}(t)}
~~|n_1, n_2 \dots \rangle \;. \label{Fermi}
\end{eqnarray}
Eq.(\ref{Fermi}) gives the correct quantum statistics of fermions: 
the states with $n_i > 1$ remain excluded as in the superalgebra case, 
because the representation imposes ${a_i^\dagger~}^2 = 0$~ 
but, because the $a_i^\dagger$'s are now even operators, we have
$[ a_i^\dagger, a_j^\dagger ] = 0$ ~and we can reach, with the correct factor,
all the allowed states.
Fermions may be thus defined as multi-mode coherent states of the $D_{1/2}$
representation of $SU(2)$ with eigenvalue 1.

\section{Anyons}
As we have the tool available, it is a simple exercise to see
what happen considering 
higher representations of $su(2)$.
To exhibit the features of the results let us consider the case of the 
coherent state of two modes in representation $D_1$: 
\begin{eqnarray}
{\rm exp}~\bigl[ a_1^\dagger + a_2^\dagger \bigr]~ |0,0\rangle \;= 
&|0,0\rangle~ + \sqrt{2} ~\bigl[
e^{i \phi_{10}(t)} \;|1,0\rangle \;+ 
~e^{i \phi_{01}(t)} ~|0,1\rangle \bigr] + ~~~ \nonumber \\
&\bigl[ e^{i \phi_{20} (t)} |2,0\rangle + 2 ~e^{i \phi_{11}(t)}|1,1\rangle + 
e^{i \phi_{02}(t)} |0,2\rangle \bigr] + ~~~   \label{Any} \\
&\sqrt{2} ~\bigl[ e^{i \phi_{21}(t)} |2,1\rangle + 
e^{i \phi_{12}(t)} |1,2\rangle \bigr] + 
e^{i \phi_{22}(t)} |2,2\rangle .~~~ \nonumber
\end{eqnarray}

We see thus the Pauli principle is no more satisfied, but we have one
generalization of it (in eq.(\ref{Any}) where $j = 1$ $n_i \le 2$, while in 
general the limit depends from the representation:
$n_i \le 2j$) of the kind introduced by Haldane~\cite{Hal} in a completely
different approach to the problem.
Let us stress that, also if this prescription gives different weights to states
(in contrast with what happen for bosons and fermions) 
the coefficients do not depend from the number of modes
as necessary to give a physical meaning to the scheme.

\section{Conclusions}

Two points have been discussed in this paper.
Standard quantum statistics for bosons and fermions have been reformulated in 
an algebraic
approach: this result is a technical one but, because whenever 
algebras have been introduced 
in some area of physics the impact has been quite sensible, 
we hope to have proposed, in this way, the framework where
quantum statistics will be described in future, in particular for the
introduction of the interaction.
The second result is that, using of the technical power of algebra, other
statistics have been studied. Boltzmann statistics is now on the same foot
of the others, the bosons have been generalized and
fermions extended to objects related to a generalization of Pauli
principle. 

Let us start from bosons. Their generalization looks truly
relevant for physics, but in great part this is illusory.
Two parameters have been introduced in eq. (\ref{ccc}), let us discuss their
meaning. $\alpha$ is the coherent 
state eigenstate and it modifies the relative weight of states with different
total number of particles $N$.
If $|\alpha| > 1$ states with low number of particles are depressed in front
of that with many particles, while $|\alpha| < 1$ has the opposite effect,
such that for $|\alpha| \ll 1$ vacuum and one-particle states are dominant.
We are thus ready to a physics completely different, as this
parameter changes. Unfortunately it is not true in standard statistical 
mechanics: in microcanonical ensemble our freedom 
does not play any role because the number of particles is fixed. Thus the same
thing must happen in gran canonical ensemble also. Explicit
calculations in the following show that $\alpha$ implies only a rescaling  
of chemical potential.

The role of the representation ({\it i.e.} the role of the parameter $k$) 
is different. To understand it let us consider states with the same $N$. 
By inspection 
$k \approx 0$ tends to concentrate all particles
in a unique state with all the others more or less empty
(it is a distribution very different from the standard ones).
$k \approx 1/2$ gives more or less flat distributions, with all states
more or less equiprobable. $k~>~1$ generates "Boltzmann-like" distributions
and, because $h(1)$ can be obtained as a contraction for $k \to \infty$ 
of $SU(1,1)$ (or, of course of ~$\widetilde{SU}(1,1)$),
for $k \gg 1$ distributions practically coincide with the Boltzmann one.

Generalized bosons thus exhibit many different behaviors when spectrum is
discrete but in continuum all of them collapse.
Indeed, while for discrete distributions the division in cells has a well
defined physical meaning, this meaning is lost and only density of states
makes sense for continuum spectrum. Eq. (\ref{ccc}) changes if we consider
levels two by two. In group theory this is realized as product of 
representations and, because we have the input of physical vacuum, this
implies to double the value of $k$. But the  $k \to \infty$ limit
is invariant under this operation.
Thus means that in the continuum limit the result is independent from the
value of $k$, {\it i.e.} it is always that obtained for $k=1/2$. 
 
To summarize, independently from $\alpha$ and $k$, the standard prediction 
are obtained in continuum and because, disregarding
first data are just arriving from new experiments on Bose condensation 
(that, we hope, in a short time, can offer some check on our predictions),
no experimental informations are known for discrete spectrum and the parameters
$\alpha$ and $k$ are up to now unphysical. 
Indeed, mimicking 
the standard statistical mechanics textbook procedure~\cite{HUA}, one 
finds first the distribution $\{ {\bar n}_i \}$ which maximizes $W\{ n_i 
\}$ with the required constraints and, at the end, from $W\{ {\bar n}_i \} $
one derives
equations of state and condensate occupation number completely equivalent to 
the usual ones, for all $k$'s and $\alpha$'s.
The effect of $k$ is nothing but to renormalize by a common factor (2$k$)
both volume $V$ and average occupation number ${\bar N}$ (this, however, in
such a way that the specific volume $v = V/{\bar N}$ is left unchanged), while
$\alpha$ affects the chemical potential, which is now given by 
\begin{eqnarray}
      \mu' = \mu + k_B T \ln |\alpha |^2 \; , \nonumber 
\end{eqnarray} 
where $\mu$ denotes the chemical potential corresponding to $\alpha =1$. 
Thus the only effect of considering $k \ne 1/2$ and/or $\alpha \ne 1$ is
to change quantities that in the continuum are not measured and unmeasurable.

From a mathematical point of view the discussion is more or less the same 
for fermions. Indeed if we gather the fermions' levels we arrive to the 
anyons (of course, as are defined in this paper) but again the dependence from 
the representation can be seen at discrete level only, because in the 
continuum limit fermions and anyons give the same result.
The point is that the physics is completely different: while for the 
statistics of bosons discrete spectrum is more or less irrelevant, most of the 
results for fermions are related to discrete spectrum.

To conclude: algebra has been introduced in quantum statistics, we could hope
that it can be as profitable as it has been in all other fields of physics.
A first result has been anyway obtained: in some sens the experimental
verification of the blackbody radiation formula has been considered, at least
unwittingly, as a confirmation of Bose's bosons; now we know that it is
only one among a lot of possible compatible scheme.

\end{document}